\documentclass[aip,jcp]{revtex4-1}
\usepackage{graphicx}
\usepackage{amsmath}
\usepackage{color}

\begin{document}

\title{Iterative derivation of effective potentials to sample the conformational space of proteins at atomistic scale}
\author{Riccardo Capelli}
\affiliation{Department of Physics, Universit\`a degli Studi di Milano, via Celoria 16, 20133 Milano, Italy}
\author{Cristina Paissoni}
\affiliation{Department of Chemistry, Universit\`a degli Studi di Milano, via Venezian 21, 20133 Milano, Italy}
\affiliation{Biomolecular NMR Unit, S. Raffaele Scientific Institute, via Olgettina 58, 20132 Milano, Italy}
\author{Pietro Sormanni}
\affiliation{Department of Chemistry, University of Cambridge, Lensfield Road, Cambridge CB2 1EW, UK}
\author{Guido Tiana}
\affiliation{Department of Physics, Universit\`a degli Studi di Milano and INFN, via Celoria 16, 20133 Milano, Italy}
\email{guido.tiana@unimi.it}
\date{\today}

\begin{abstract}
The current capacity of computers makes it possible to perform simulations of small systems with portable, explicit-solvent potentials achieving high degree of accuracy. However, simplified models must be employed to exploit the behaviour of large systems or to perform systematic scans of smaller systems. While powerful algorithms are available to facilitate the sampling of the conformational space, successful applications of such models are hindered by the availability of simple enough potentials able to satisfactorily reproduce known properties of the system. 
We develop an interatomic potential to account for a number of properties of proteins in a computationally economic way. The potential is defined within an all-atom, implicit solvent model by contact functions between the different atom types. The associated numerical values can be optimised by an iterative Monte Carlo scheme on any available experimental data, provided that they are expressible as thermal averages of some conformational properties.
We test this model on three different proteins, for which we also perform a scan of all possible point mutations with explicit conformational sampling. The resulting models, optimised solely on a subset of native distances, not only reproduce the native conformations within a few Angstroms from the experimental ones, but show the cooperative transition between native and denatured state and correctly predict the measured free--energy changes associated with point mutations. Moreover, differently from other structure-based models, our method leaves a residual degree of frustration, which is known to be present in protein molecules.
\end{abstract}

\maketitle

\section{Introduction}

The development of new algorithms and of powerful computers currently allows to study in explicit solvent large conformational changes of small proteins and peptides\cite{Piana:2013jy,Bowman:2011hm} and smaller conformational changes in large proteins\cite{Sutto:2013gy}, but still with some computational effort. More complex simulations, like those of large changes in large systems, of aggregation of many protein chains, or of systematic mutation scans still require the use of models with simplified degrees of freedom. Pasrticularly useful in this respect are implicit-solvent models controlled by simple potentials, like those involving only contact functions (and thus not requiring the lengthy calculation of accessible surface areas).

While a thorough sampling of the conformational space of a protein system described by such simplified models is now rather affordable even for large proteins and even describing explicitely all the heavy atoms of the system, the determination of a simple potential capable of recapitulate the properties of a protein is still a challenging problem. The basic requirement for such a potential is to make the native conformation of proteins stable, as entailed by the thermodynamic hypotesis \cite{Anfinesn:1973vt}. Several different approaches were used to implement this requirement. Using associative-memory potentials \cite{Goldstein:1992tt} which encodes for correlation between protein sequence and native structure, motivated by the theory of neural networks, it was possible to predict the native conformation of a number of proteins from the knowledge of their sequence \cite{Prentiss:2006uv}, even if in a simplified geometry. Minimizing simulateneously the potential in the native conformation of several proteins with respect to their competitive conformations \cite{Mirny:1996hz} allowed to design a potential capable of identifying the native conformation within the framework of a minimal model of protein-like polymers. However, this potential failed to distinguish the native from alternative conformations in the case of real proteins, mainly because the exploration of competitive conformations was computationally too demanding. Similar approaches were carried out sampling competitive conformations with a Monte Carlo algorithm within a bead model \cite{Hao:1996vq}, or through a variational approach\cite{Seno:1998wh,Seno:1998tw}. Anyway, they have not been completely succesful for real proteins, always yielding poor results for full-sized protein molecules. Simple potentials originally designed for structure prediction have also been succesfully used to sample non-native states of small proteins \cite{Shirvanyants:2012,Kimura:2014}.

A simpler approach is to use structure-based models, specific for each protein. In this case one desists from building a universal potential, capable of predicting the native conformation from its sequence, and focuses on the investigation of the properties of a protein of known structure. This is the case, for example, of the popular Go model \cite{Go:1983in}, which is a direct implementation of the principle of minimal frustration \cite{Bryngelson:1987uu}. For example, with a coarse-grained Go model it was possible to simulate the cotranslational folding of 100-residues proteins within the whole, explicitly-represented ribosome \cite{Elcock:2006ke}.

Structure-based models have been succesful in reproducing a number of features of proteins, expecially related to the native state and to the transition between the native and the denatured state \cite{Clementi:2008vl,Hills:2009bt,Wolff:2011km}. However, they are not able to describe properly non-native interactions, and consequently cannot account for the properties of the denatured state, for intermediate states stabilized by non-native interaction, for protein aggregation, and for all those properties that emerge from the competition between native and non-native interactions.

In the present work we build an implicit-solvent, model which describes all heavy atoms and which retains the computational handiness of minimally-frustrated models, but do not suffer their limitations. A key feature of this model is that it must fold to the native conformation of the protein.  For this purpose, we develop a strategy to design a potential between the different atom types  to make the  equilibrium state of the model at low temperature unique and equal to the native conformation. 

Once the potential has been designed, one can sample the conformational space of the system, and thus study thermodynamic quantities other than those used as input in the design algorithm. In this way it is possible to understand what properties of the protein are a necessary consequence of the stability of its native state. We show that the low-temperature equilibrium state is unique and is identical to the experimental  the two-state character of the transition between native and denatured state, the energetic effect of experimentally-characterized mutations and some features of the denatured state can be reproduced without any further input to the system.

The potential is chosen as the sum of two-body terms, accounting for the interactions between pairs of atoms, and shaped as a double spherical well. This choice allows a remarkably fast sampling of the conformational space of the protein system by means of Monte Carlo (MC) algorithms. Each two-body term is determined by a single parameter which determines the depth of the energy wells and which depends of the chemical species involved. Operatively, the set of energy parameters associated with all pairs of chemical species are optimized according to an iterative Monte Carlo algorithm, employing the reweighting scheme developed by Norgaard and coworkers \cite{Norgaard:2008ep}, to make the thermal averages of a set of inter-atomic distances match the value they display in the experimental native conformation of the protein. A sequence-dependent potential on the backbone dihedrals is also introduced to favour the formation of secondary structures. 
The resulting potential will be minimally frustrated if this is required by the system to display a stable native state, but this ingredient is not pushed by hand. In fact, atoms of the same type but belonging to different positions along the chain interact in the same way, and consequently can stabilize, even strongly, non-native interactions.  

\section{The model and the optimization of the potential}

In the model we developed proteins are described through all their heavy atoms. All bond distances, backbone angles and dihedrals of the peptidic bond are mantained rigidly fixed, corresponding to their experimental values. The Ramachandran dihedrals can move freely, while the residue can move among the rotamers defined in ref. \cite{Lovell:2000ui}. 

Starting from the knowledge of the native conformation of the protein, the potential which controls it has the form
\begin{equation}
U=\sum_{i<j}U_{ij}+U_{dih}
\label{eq:potential}
\end{equation}
The former, two--body term is a two-well spherical potential which depends on the positions $r_i$ and on the kind $\sigma_i$ of the atoms involved, in the form
\begin{equation}
U_{ij}=
\begin{cases} 
+\infty & \text{if $|r_i-r_j|<r_{HC}(\sigma_i,\sigma_j)$} \\
B_{\sigma i\sigma j} & \text{if $r_{HC}(\sigma_i,\sigma_j)<|r_i-r_j|<r_s(\sigma_i,\sigma_j)$} \\
B_{\sigma i\sigma j}/2 & \text{if $r_s(\sigma_i,\sigma_j)<|r_i-r_j|<r_m(\sigma_i,\sigma_j)$} \\
0 & \text{if $|r_i-r_j|>r_m(\sigma_i,\sigma_j)$}.
\end{cases}
 \end{equation}
Atoms that are separated by less than 9 other atoms along the backbone do not display attractive two--body interactions. The minimum of the well has energy $B_{\sigma,\pi}$ that depends on the types $\sigma$,$\pi$ of the atoms involved. Different atoms in different amino acids are regarded as different atom types , giving a total of 163 atom types. Defining a native contact between two atoms if the two atoms are closer than $d_{th}=3.8\AA$, we label $d'_N(\sigma,\pi)$ the maximum distance between atoms of kind $\sigma$ and $\pi$ in all native contacts of the protein. The hard-core radius for that pair of atom types is then defined as $r_{HC}(\sigma,\pi)  = 0.67\, d'_N(\sigma,\pi)$, at the radius $r_s(\sigma,\pi) = 0.78 \,d'_N(\sigma,\pi)$ the energy depth of the well is decreased by a factor 2, and the overall interaction range is $r_m(\sigma,\pi) = 1.4\, d'_N(A,B)$. 

The potential on the Ranachandran dihedrals $\{\phi_i\}$ and $\{\psi_i\}$ is meant to account for the interactions between atoms close along the chain, and thus to induce the formation of local secondary structure. It has the form 
\begin{align}
U_{dih}&=\sum_i \left[ \frac{\epsilon_\alpha k_{\alpha i}}{\sigma_{\phi\alpha}} e^{-(\phi_i-\phi_{0\alpha})^2/2\sigma_{\alpha}^2} +  \frac{\epsilon_\beta k_{\beta i}}{\sigma_{\phi\beta}} e^{-(\phi_i-\phi_{0\beta})^2/2\sigma_{\phi\beta}^2} +  \right. \nonumber\\
&+\left.  \frac{\epsilon_\alpha k_{\alpha i}}{\sigma_{\psi\alpha}} e^{-(\psi_i-\psi_{0\alpha})^2/2\sigma_{\psi\alpha}^2} +  \frac{\epsilon_\beta k_{\beta i}}{\sigma_{\psi\beta}} e^{-(\psi_i-\psi_{0\beta})^2/2\sigma_{\psi\beta}^2}  \right],
\end{align}
where $\epsilon_\alpha,\epsilon_\beta<0$ are the energy constants that set the weight of the dihedral potential with respect to the two--body potential and to each other and are chosen as $\epsilon_\alpha =-80$ and  $\epsilon_\beta=-200$ in order to allow the formation of secondary structure at $T\sim 1$  but, at the same time, not to make the two--body potential irrelevant\cite{suppmat}. The quantities $\phi_{0\alpha}=-57^\circ$, $\psi_{0\alpha}=-47^\circ$, $\phi_{0\beta}=-129^\circ$ and $\psi_{0\beta}=-124^\circ$ are the averages of Ramachandran dihedrals in typical $\alpha$ and $\beta$ conformations, respectively, while the quantities $\sigma_{\phi\alpha}=25^\circ$, $\sigma_{\psi\alpha}=30^\circ$, $\sigma_{\phi\beta}=30^\circ$ and $\sigma_{\psi\beta}=35^\circ$ are the associated standard deviations (see \cite{suppmat}).  The quantities $\{k_{\alpha i}\}$ and $\{k_{\beta i}\}$ are the sequence--dependent propensities for the $i$th amino acid of $\alpha$ and $\beta$ structure, respectively, calculated with PSIPRED\cite{Jones:1999}. We choose not to make it dependent on the specific native conformation not to bias the formation of secondary structures which could be stabilized by tertiary contacts. The dihedral potential is not affected by the optimization procedure.

Before starting the simulation, for each protein a set of $n_r=100$ pairs of atoms $(i_K,j_K)$ are selected in such a way that they do not belong to amino acids closer than 4 along the sequence, and the distances $d_N(i_K,j_K)$ between each pair in the native conformation recorded. This choice guarantees that the implementation of all the $d_N(i_K,j_K)$ in a  conformation of the protein makes it identical to the native conformation, with an RMSD smaller than 1\AA. The whole idea is to optimize the interaction matrix $B_{\sigma,\pi}$ so that the thermal average of the distance between each pair of atoms $i_K$ and $j_K$ is equal to the distance $d_N(i_K,j_K)$ they have in the experimental native conformation, that is
\begin{equation}
\langle |r_{iK}-r_{jK}|\rangle = d_N(i_K,j_K) \;\;\text{for each $K$ with $1\leq K \leq n_r$},
\end{equation}
and consequently that the equilibrium conformation of the protein is the native one.

To implement this idea, we start from an interaction matrix in which $B_{\sigma\pi}=-0.5$ if there is al least one pair of atoms $\sigma$ and $\pi$  such that $d'_N(\sigma,\pi)<d_{th}$ and 0 otherwise. The choice of the initial matrix is not really critical. Making use of the potential (\ref{eq:potential}), a MC sampling is carried out and a set of conformations at temperature $T=1$, which is regarded as reference temperature and sets the energy units (Boltzmann's constant is also set to 1), is recorded.  At the end of the MC sampling, the average distances $\langle |r_{iK}-r_{jK}|\rangle$ are calculated from the recorded conformations and the  $\chi^2$ between them and the native distances $d_N(i_K,j_K)$ is evaluated, using $0.4$\AA as error allowed for all contacts in the definition of $\chi^2$. The  $B_{\sigma\pi}$ are optimized to minimize the  $\chi^2$ making use of a zero-temperature random minimization. At each step of the minimization, the average distances $\langle |r_{iK}-r_{jK}|\rangle'$ according to the modified potential $U'$ are calculated following the reweigting scheme described in ref. \cite{Norgaard:2008ep}, that is
\begin{equation}
\langle |r_{iK}-r_{jK}|\rangle '=\frac{1}{Z}\sum_t   |r_{iK}(t)-r_{jK}(t)|\cdot\exp\left[\frac{-U'(t)+U(t)}{T}\right],
\end{equation}
where 
\begin{equation}
Z=\sum_t \exp\left[\frac{-U'(t)+U(t)}{T}\right]
\end{equation}
and the index $t$ runs over 5000 conformations recorded during the MC sampling carried out with the potential $U$. Then, a new MC simulation is carried out with the new potentials and the procedure is repeated iteratively 100 times.

The MC sampling is carried out with a parallel-tempering \cite{Swendsen:1986} scheme. The MC moves are pivots on the backbone dihedrals, combinations of pivots on adjacent backbone dihedrals \cite{Shimada:2002ij} to produce local moves, and discrete moves of the side chains among all possible rotamers. In each simulation 8 replicas of the system are used, at temperatures ranging from 1 to 1.75. Each MC iteration is carried out for $10^7$ steps for each replica. Every $10^3$  steps after the half of the simulation the conformation belonging to the replica at $T=1$ is recorded. More details about the model and the optimization scheme are given in \cite{suppmat}.
 
Important questions concerning the optimization procedure are whether the optimal potential is unique and to which extent it is portable among different proteins. A comparison of two interaction matrices for protein G, optimized independently on each other, give a correlation coefficient of 0.74, with matrix elements more similar towards the ends of the distribution and more dissimilar towards zero\cite{suppmat}. This suggests that the most stabilizing matrix elements are rather independent on the realization of the optimization procedure, but depends only on the protein. On the other hand, the correlation between the matrix elements associated with the same atom types in two proteins, specifically protein G and villin\cite{suppmat}, is 0.08, indicating that the optimized potential is not portable among proteins.

 \section{Folding of villin headpiece, GB1 domain and src--SH3}
 
A necessary condition that the optimized models have to satisfy is to display the experimental native conformation as low--temperature equilibrium state. Although the optimization was carried out towards the native distances, it is not straightforward that this is enough to let the model satisfy such a necessary condition. In the present model the interaction between two atoms depend on their kind, not on their position in the protein. This introduces frustration\cite{Toulouse:1977} in the system as, differently from the Go models\cite{Go:1983in}, the optimal interaction matrix is not simply that in which two atoms strongly attract each other if they are in contact in the experimentally-determined native conformation. The model satisfies the above necessary conditions if the optimization procedure is able to lower the energy of the native conformation below that of the competing conformations\cite{Shakhnovich:1993vz} or, in other words, if it can minimize its degree of frustration\cite{Bryngelson:1987uu}.
 
\begin{figure}
\includegraphics[width=\linewidth]{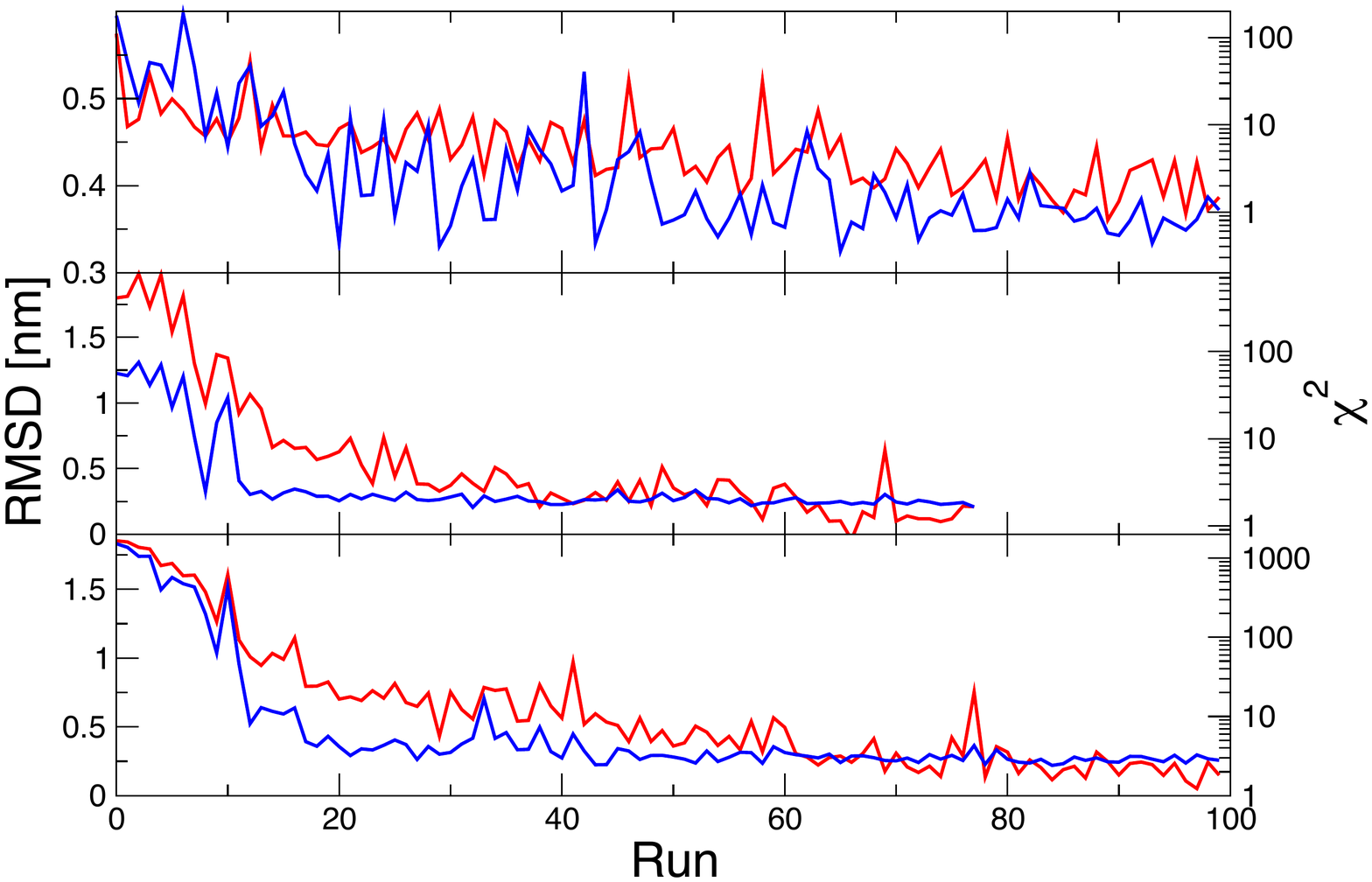}
\caption{The $\chi^2$ between average and native distances (red curve, in semi--log scale) and the average RMSD to the experimental native conformation (blue curve) as a function of the number of iterations of the MC sampling for villin (upper panel), protein G (middle panel) and SH3 (lower panel).}
\label{fig:chi2}
\end{figure}

We have tested the model on three widely--studied proteins. These are the villin headpiece (pdb code 1VII), the B1 domain of protein G (pdb code 1PGB) and the Sh3 domain of Src (pdb code 1FMK). The optimization procedure is illustrated in Fig. \ref{fig:chi2}, where the $\chi^2$ to the set of native distances and the average RMSD to the native conformation is displayed as a function of the number of iterations. Each iteration consists of a MC sampling and an optimization of the interaction matrix. In the case of protein G and SH3 there is a sharp drop of both $\chi^2$ and average RMSD in the first 20 iterations. Protein G reaches a stationary $\chi^2\approx 1$ and an average RMSD $\approx 0.3$ nm, while SH3 reaches  $\chi^2\approx 2$ and an average RMSD $\approx 0.3$ nm. Interestingly, while, the average RMSD reaches it stationary value around the 20th iteration and remains stationary since then, the $\chi^2$ takes a longer time to find its minimum, indicating that RMSD does not capture completely all structural features of the native state. The behavior of villin is more noisy, most likely because its size is smaller than that of the other two proteins. Anyway, it can converge to $\chi^2\approx 2$ and $\langle \text{RMSD}\rangle\approx 0.4$ nm after 100 iterations.
 
\begin{figure}
\includegraphics[width=\linewidth]{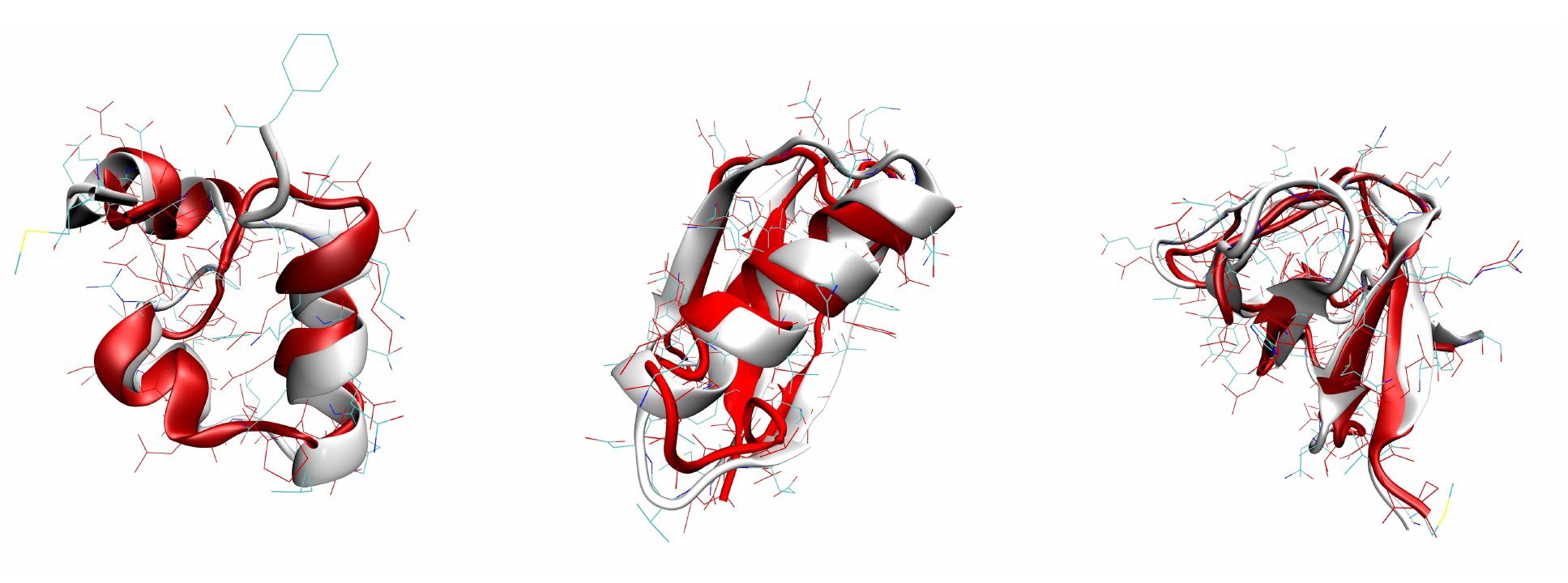}
\caption{Comparison between experimental (white) and simulated (red) 3D structure for (a) C-terminal chicken villin headpiece domain, (b) B1 domain of streptococcal protein G, (c) human src SH3 domain.}
\label{fig:native}
\end{figure}

The minimum--energy conformations found with the interaction matrix obtained in the last iteration of the optimization process is displayed in Figure \ref{fig:native} for each of the three proteins. The RMSD to the experimental native conformations are $0.41$ nm for villin, $0.14$ nm for protein G and $0.18$ nm for SH3. No low--energy conformations with RMSD markedly larger than these are observed\cite{suppmat}.

\begin{figure}
\includegraphics[width=\linewidth]{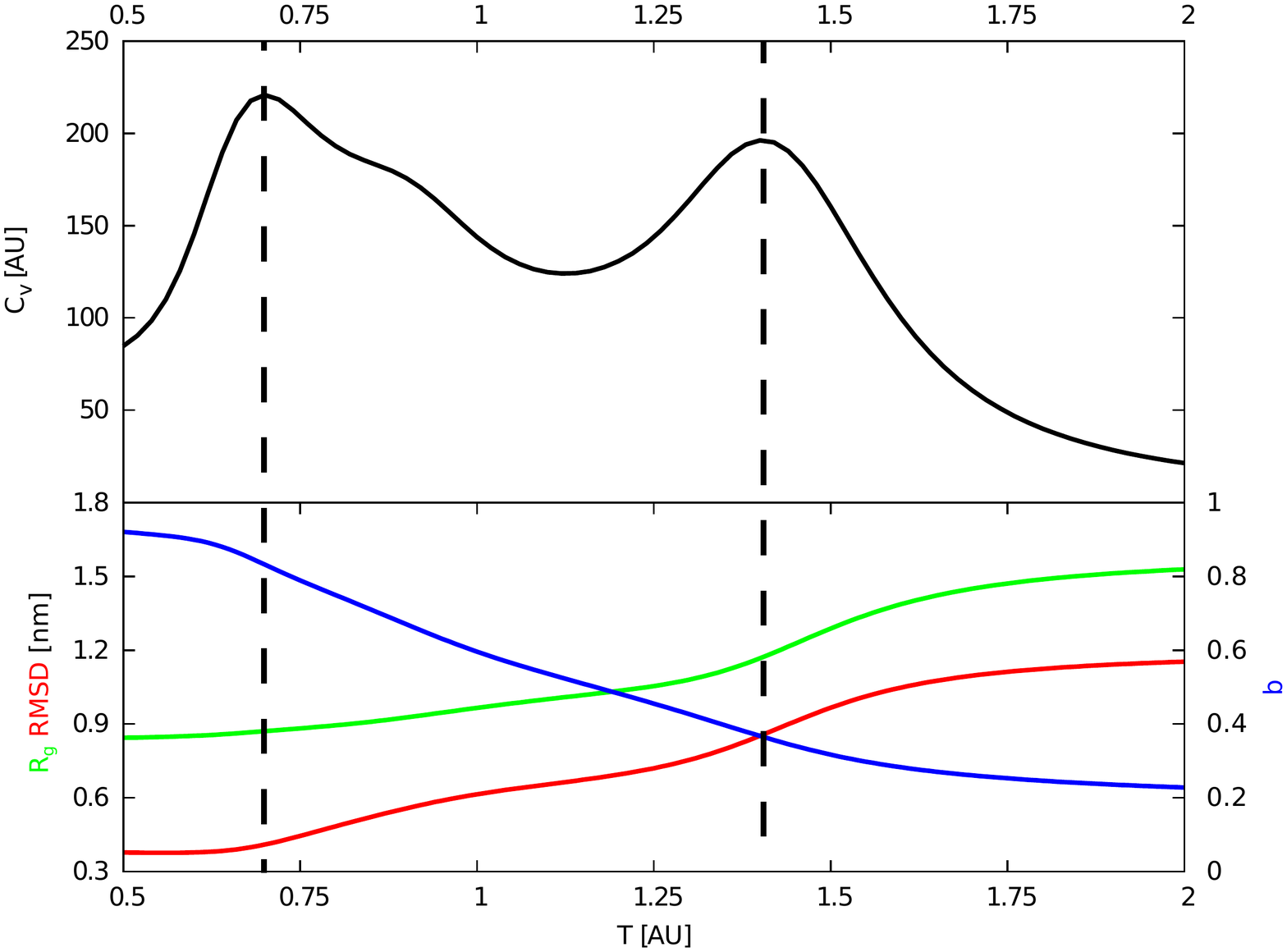}
\caption{The specific heat (above), average RMSD (in red, below), gyration radius (in green) and fraction $q$ of native contacts (in blue) as a function of temperature for villin.}
\label{fig:cv_villin}
\end{figure}

\begin{figure}
\includegraphics[width=\linewidth]{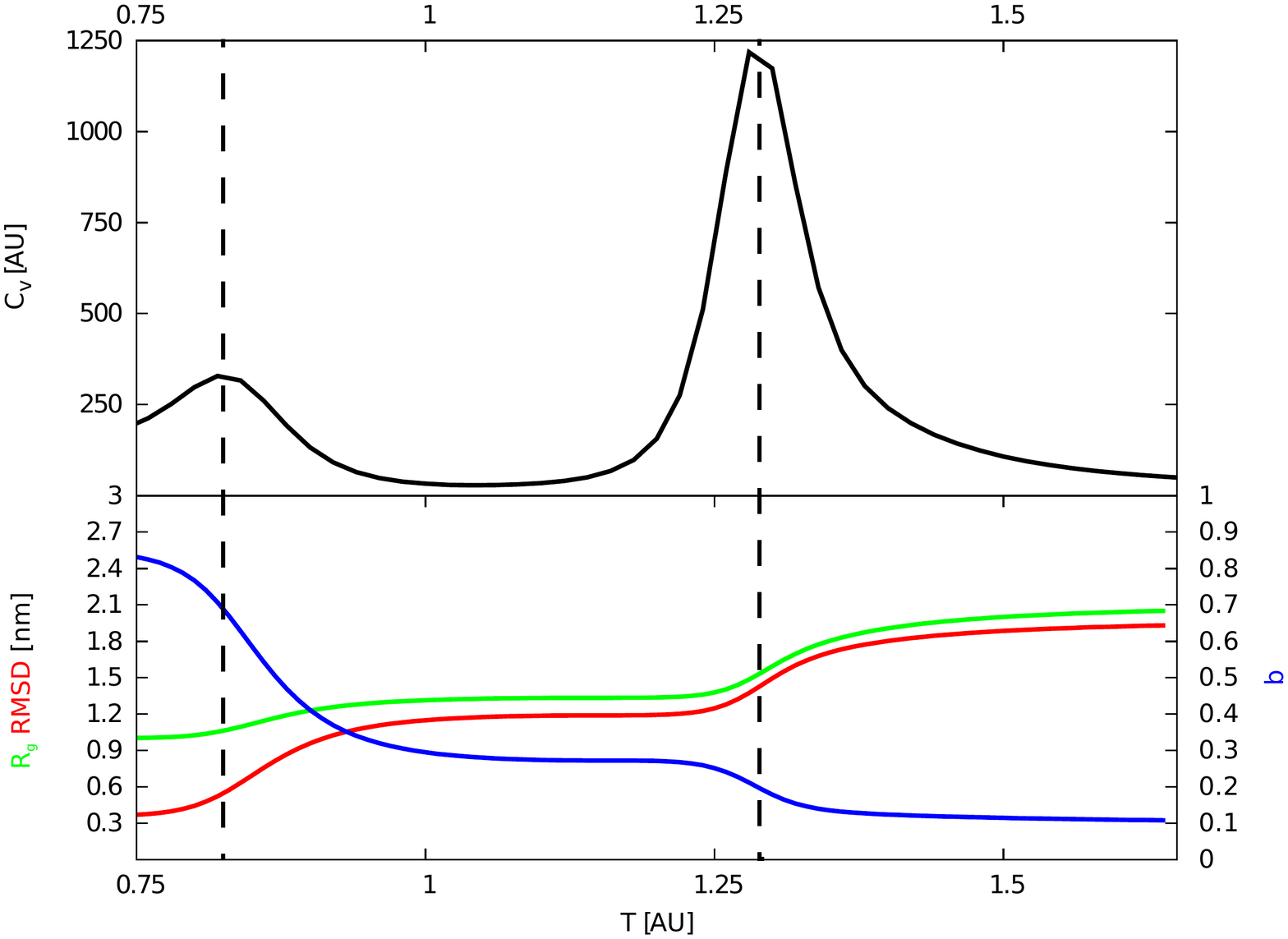}
\caption{The specific heat (above), average RMSD (in red, below), gyration radius (in green) and fraction $q$ of native contacts (in blue) as a function of temperature for protein G.}
\label{fig:cv_protG}
\end{figure}

\begin{figure}
\includegraphics[width=\linewidth]{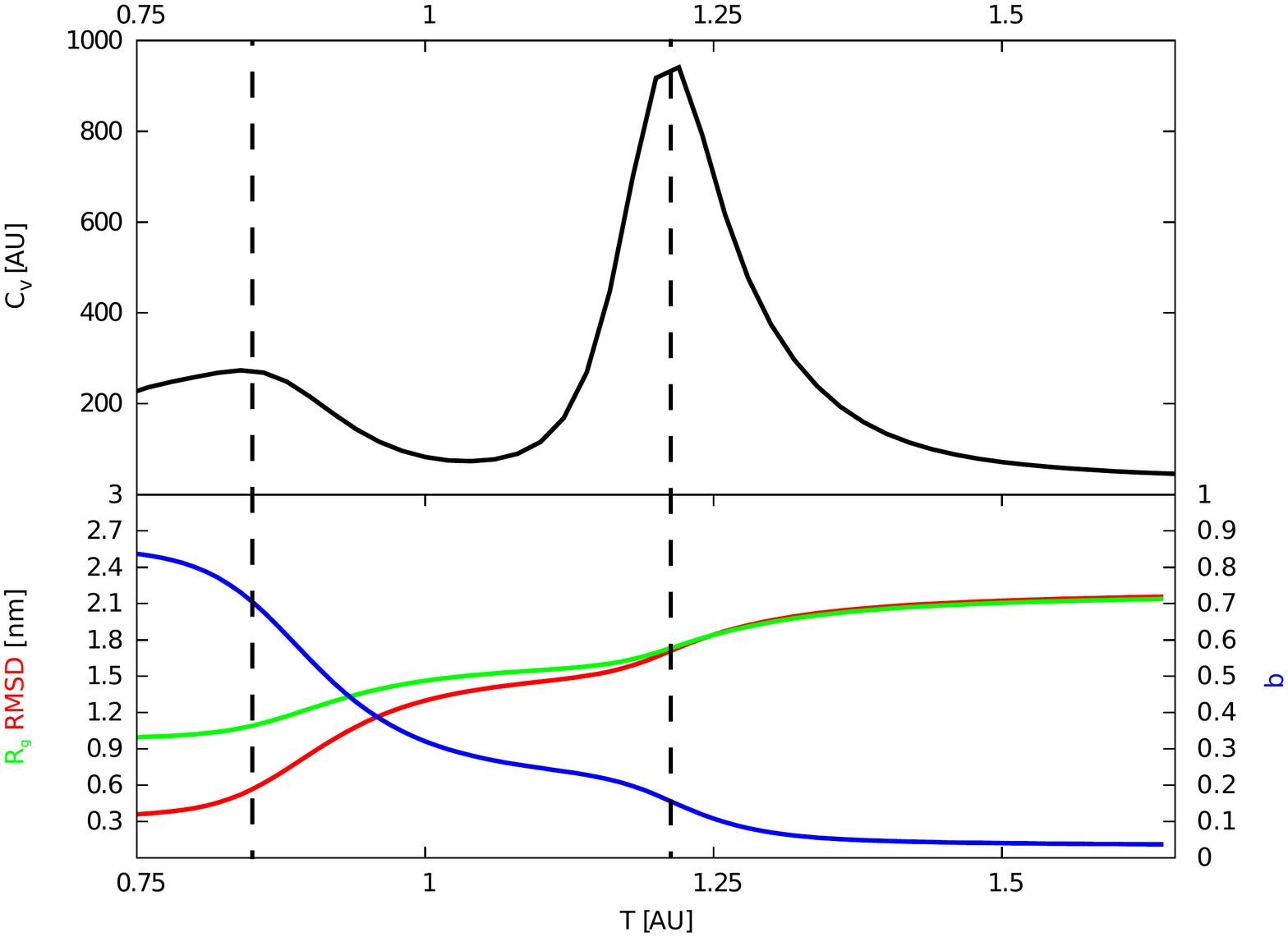}
\caption{The specific heat (above), average RMSD (in red, below), gyration radius (in green) and fraction $q$ of native contacts (in blue) as a function of temperature for SH3.}
\label{fig:cv_SH3}
\end{figure}

The thermodynamic properties of the three proteins as a function of temperature, calculated with a weighted--histogram algorithm \cite{Ferrenberg:1989tf}, are summarized in Figs. \ref{fig:cv_villin}, \ref{fig:cv_protG} and \ref{fig:cv_SH3}, respectively. All of them display two peaks in the specific heat. The lower--temperature one (at $T_f=0.68$ for villin, $T_f=0.82$ for protein G and $T_f=0.84$ for SH3) marks the folding transition, as testified by the change and in average RMSD (calculated on all heavy atoms) and fraction of native contacts $q$ that takes place at those temperatures. The higher--temperature peak ($T_{cg}=1.39$ for villin, $T_{cg}=1.29$ for protein G and $T_{cg}=1.21$ for SH3) corresponds to the coil--globule transition (cf. the change in average gyration radius at those temperatures).

In agreement with the experimental findings, and not unexpectedly because of their difference in size, villin results less stable than protein G (folding temperatures at neutral pH are $73.5^\circ$C for villin\cite{GodoyRuiz:2008jf} and $87.5^\circ$C for protein G\cite{Alexander:1992um}), and the folding transition less cooperative. In fact, the ratio $\kappa$ between calorimetric and van't Hoff enthalpy, which takes its minimum value of 1 for a pure two--state transition\cite{Pri:1974}, results from model calculations to be $\kappa=5.41$ for villin and $\kappa=1.89$ for protein G, to be compared with the experimental values  $\kappa=4.52$ for villin\cite{GodoyRuiz:2008jf} and $\kappa=1.07$ for protein G\cite{Alexander:1992um}, while it is $\kappa=3.12$ for our model of SH3. However, in all cases the model understimate the two--body character of the folding transition, as already observed for other models which only include two--body interactions\cite{Chan:2000wk}.

It should be noted that the model displays a folding transition for all the three proteins at temperatures lower than 1, that is the temperature at which the interaction potential has been optimized to reproduce the native distances. This suggests that the computational limitations in the optimization of the interaction matrix result not much in errors in the conformational properties of low--temperature states, but in a decreased thermodynamic stability. 
 
The free energies of the three proteins as a function of the RMSD and of the gyration radius, calculated with a weighted--histogram algorithm  \cite{Ferrenberg:1989tf}, are displayed in Fig. \ref{fig:free} in the case of a temperature below the folding transition, a temperature between the folding and the coil--globule transition and a temperature above the coil--globule transition. For none of the proteins the free energy profile highlights detectable intermediates. The globular denatured state (at $T=1.0$) is in all cases rather native--like, displaying RMSD of the order of 0.5--0.6 nm.

\begin{figure}
\includegraphics[width=10cm]{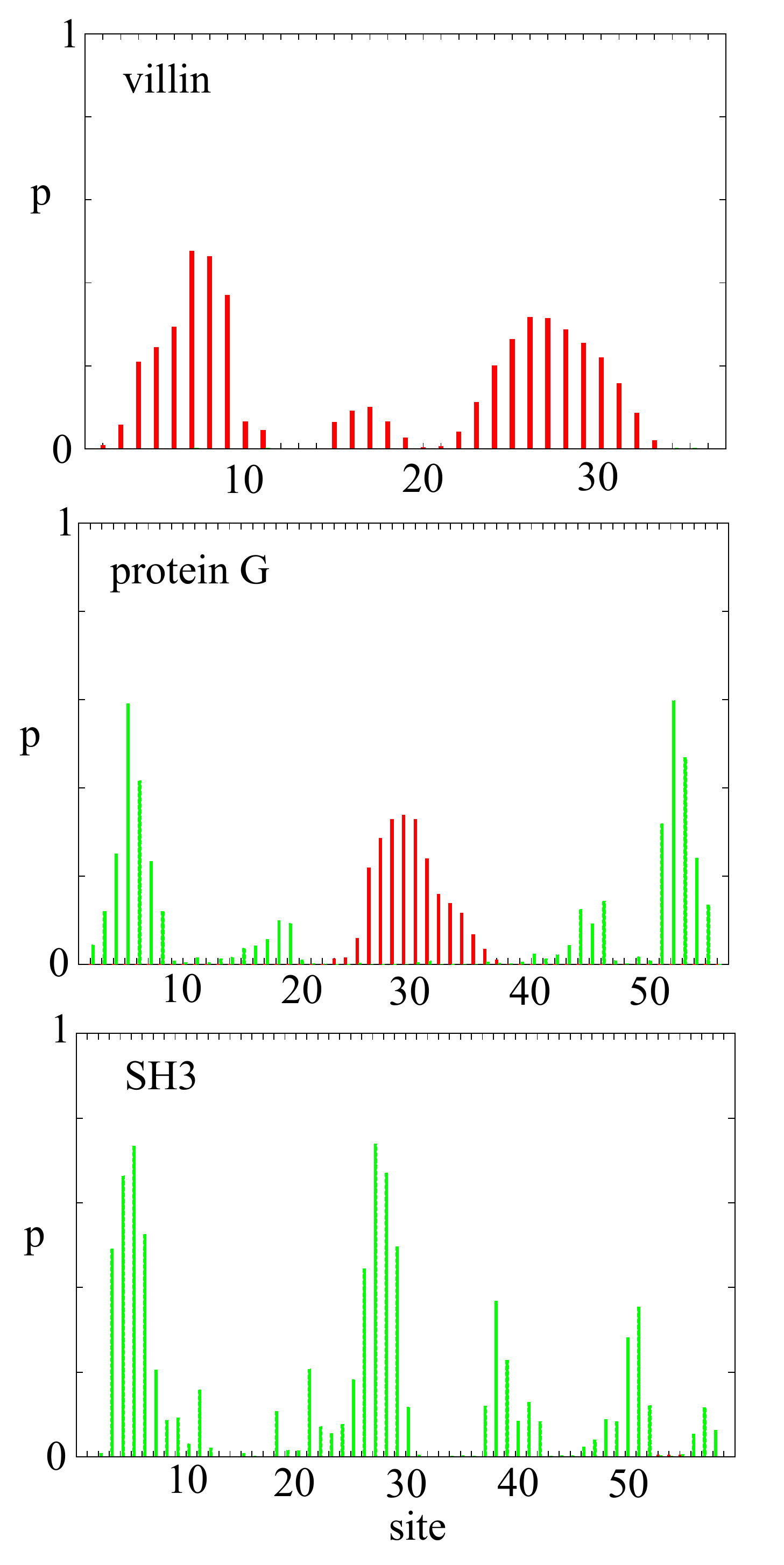}
\caption{The formation probability of alpha--helices (in red) and beta--strands (in green) in the denatured state ($T=1.0$) of villin, protein G and SH3.}
\label{fig:stride}
\end{figure}

The reason for such a low RMSD is the formation of residual, largely native--like, structure in the denaturated state, as shown in Fig. \ref{fig:stride}. In the case of villin, residual alpha--helical structure is larger in the N--terminal segment, slightly smaller in the C-terminal segment, and marginal in the central segment. These ratios are in agreement with circular--dochroism spectra of isolated fragments of villin\cite{Tang:2004ik} and with explicit--solvent molecular--dynamics simulations\cite{Wickstrom:2006fm}. The denatured state of protein G displays in native--like residual structure in the two hairpins and in the helix, but not the non--native turns  osberved in the acid--denatured state by NMR\cite{Sari:2000ua}.  The denaturated state of SH3 is enriched in beta-starnd structure, a feature that is not observed in NMR experiments with urea, which indicates abundance of non-native helices\cite{Rosner:2010kg}. There can be two straightforward reasons for this dicrepancy. First, our model simulates a thermal--denatured state, while in NMR experiments the protein is destabilized by urea. Moreover, while the agreement with experiments of the other two proteins concerns native--like structure, in the case of SH3 the model is not able to predict non--native residual structure. This could be due to the fact that the optimization of the potential to stabilize the native conformation over--minimize the frustration of the system. However, a nice feature of the present approach is that, in principle, one can optimize the interaction matrix to reproduce the native distances at low temperature and, simultaneously, the data observed in the denatured state at higher temperature.
 
 \section{Stability of GB1 domain and src--SH3 against mutations}

In the case of protein G and SH3, the free--energy changes $\Delta\Delta G_{UN}$ of the native state upon mutation was measured for a large number of mutations\cite{McCallister:2000do,Grantcharova:1998vw}.  Within the present model, the relatively small computational cost of sampling the conformational space allows to simulate the effect of each mutation, and compare the result with the experimental data. These simulations have two goals. First, the comparison between experimental and calculated $\Delta\Delta G_{UN}$ can contribute to validate the model. Moreover, the simulation has access to conformational properties of the mutated system that cannot be studied experimentally in a direct way.

\begin{figure}
\includegraphics[width=\linewidth]{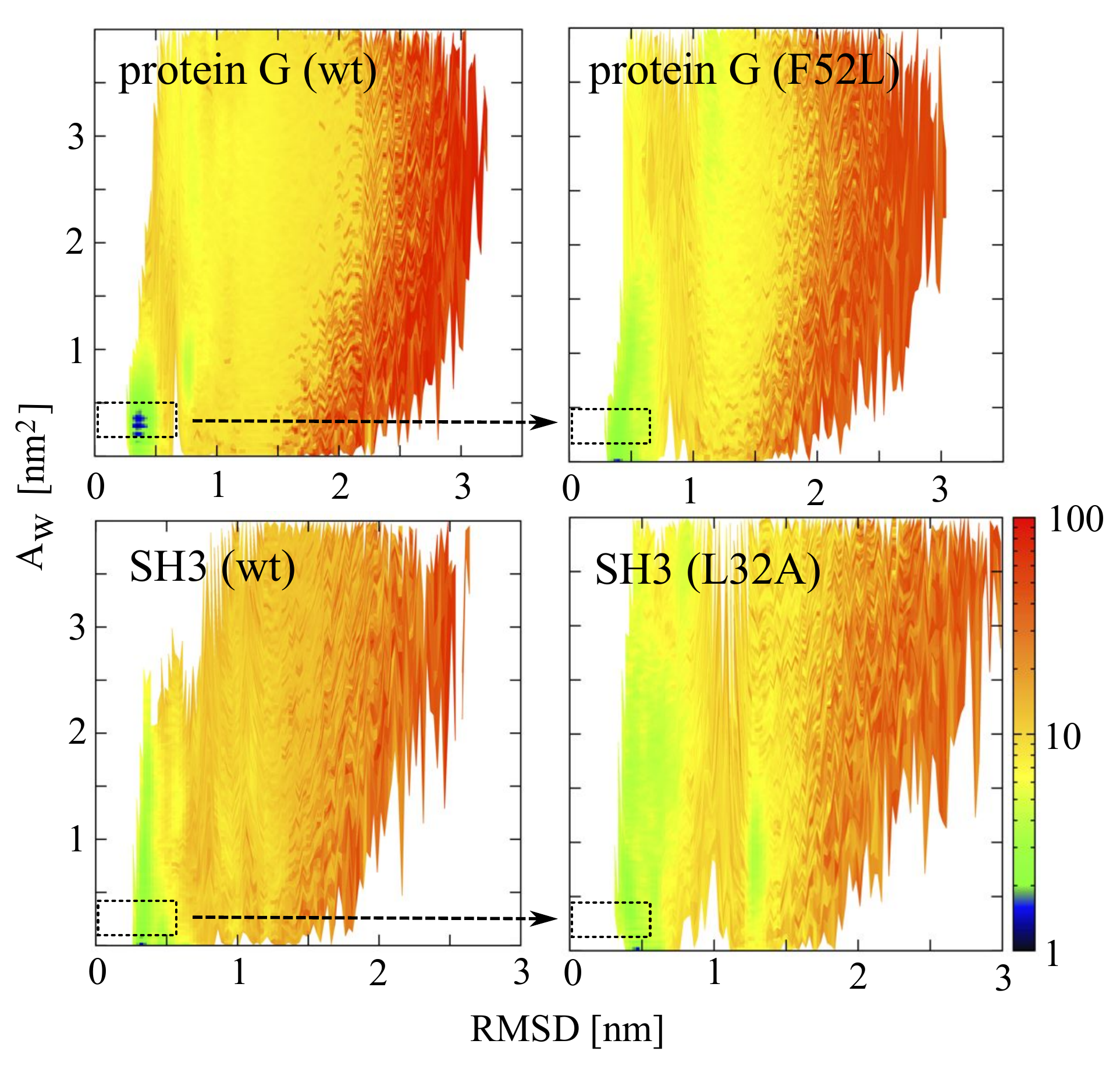}
\caption{The comparison of the  free--energy profiles of protein G (above) and SH3 (below) for the wild--type sequence (left) and an example of mutation (right). The dashed rectangle identifies the native state}
\label{fig:freemut}
\end{figure}

Operatively, a mutation means changing the atom types of the mutated residue, which interact with the same matrix elements of the wild--type protein (no further optimization is carried out), and updating the secondary--structure propensities in the dihedral potential.  For each of the mutation reported for protein G\cite{McCallister:2000do} and SH3\cite{Grantcharova:1998vw} we have carried out an equilibrium simulation,  obtaining  the free--energy profile of the wild--type  ($F_{wt}$) and of the mutated ($F_{mut}$) protein, as a function of RMSD and exposed area\cite{Eisenhaber:1995} $A_W$ of the tryptophanes. The reason for the choice of $A_W$ is that experimental $\Delta\Delta G_{UN}$ were obtained from kinetic experiments in which the measured quantity is the fluorescence of the tryptophanes, which depend on their molecular environment. From these free energies, we have calculated the free--energy differences in a two--state approximation, that is
\begin{equation}
\Delta\Delta G_{UN}=-T\log\frac{ p_N^{wt}(1-p_N^{mut})}{ p_N^{mut}(1-p_N^{wt}) },  
\end{equation}
where
\begin{align}
p_N^{wt}&\equiv\int_{\cal N}d\,\text{RMSD}\,dA_W\;\exp[-F^{wt}(\text{RMSD},A_W)/T] \nonumber\\
p_N^{mut}&\equiv\int_{\cal N}d\,\text{RMSD}\,dA_W\;\exp[-F^{mut}(\text{RMSD},A_W)/T] 
\end{align}
and the native region ${\cal N}$ in the free--energy profiles is that defined in Fig. \ref{fig:freemut}. 

\begin{figure}
\includegraphics[width=\linewidth]{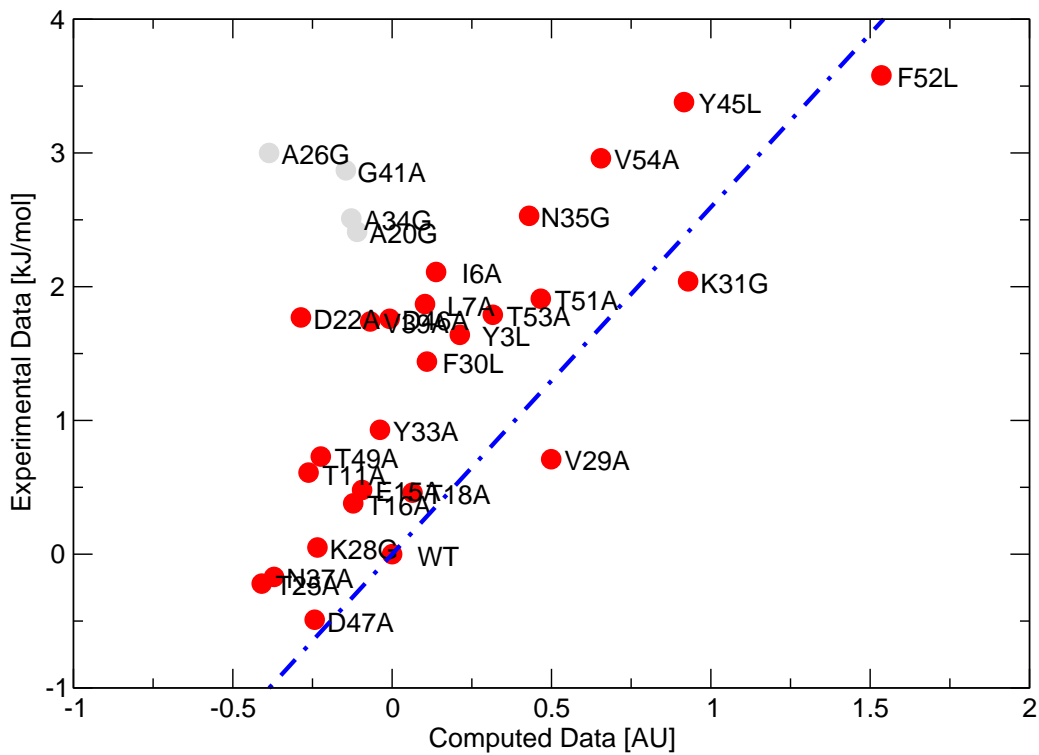}
\caption{A comparison of the predicted with the experimental $\Delta\Delta G$ of all the mutations measured for protein G. Grey points mark outliers, while the blue line is the best linear fit. The correlation coefficient is 0.79.}
\label{fig:mut_G}
\end{figure}

\begin{figure}
\includegraphics[width=\linewidth]{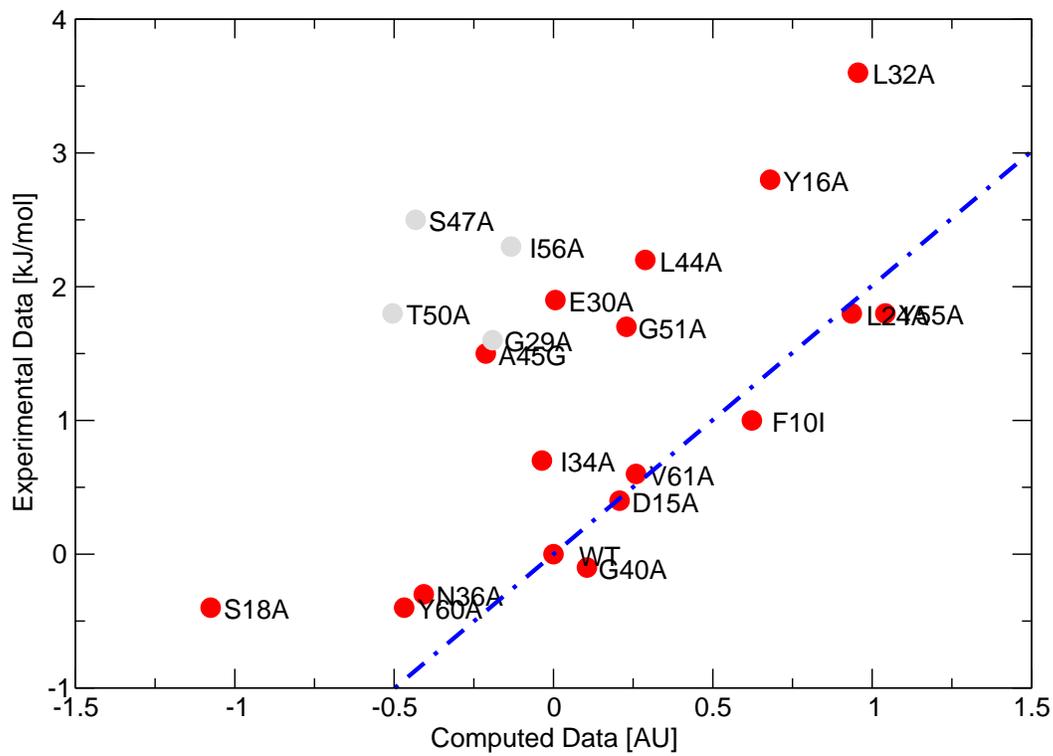}
\caption{Same as Fig. \protect\ref{fig:mut_G} for SH3. The correlation coefficient is 0.73.}
\label{fig:mut_SH3}
\end{figure}

The comparison between experimental and computed $\Delta\Delta G_{UN}$ is displayed in Figs. \ref{fig:mut_G} and \ref{fig:mut_SH3} for protein G and SH3, respectively. The correlation coefficients are, respectively, 0.57 and 0.50, which increase, respectively, to 0.79 and 0.73 if we exclude four outliers. These values correspond to the optimal choice of the native region ${\cal N}$. Interestingly, such outliers correspond to sites which display in the calculations large native--like structure or does not display the non--native secondary structures measured by NMR\cite{Sari:2000ua,Rosner:2010kg}. Consequently, one could make the hypothesis that the poor agreement between theoretical and experimental $\Delta\Delta G_{UN}$ is associated with the overstimation of native structure in the denatured state discussed in the previous Section.

\begin{figure}
\includegraphics[width=10cm]{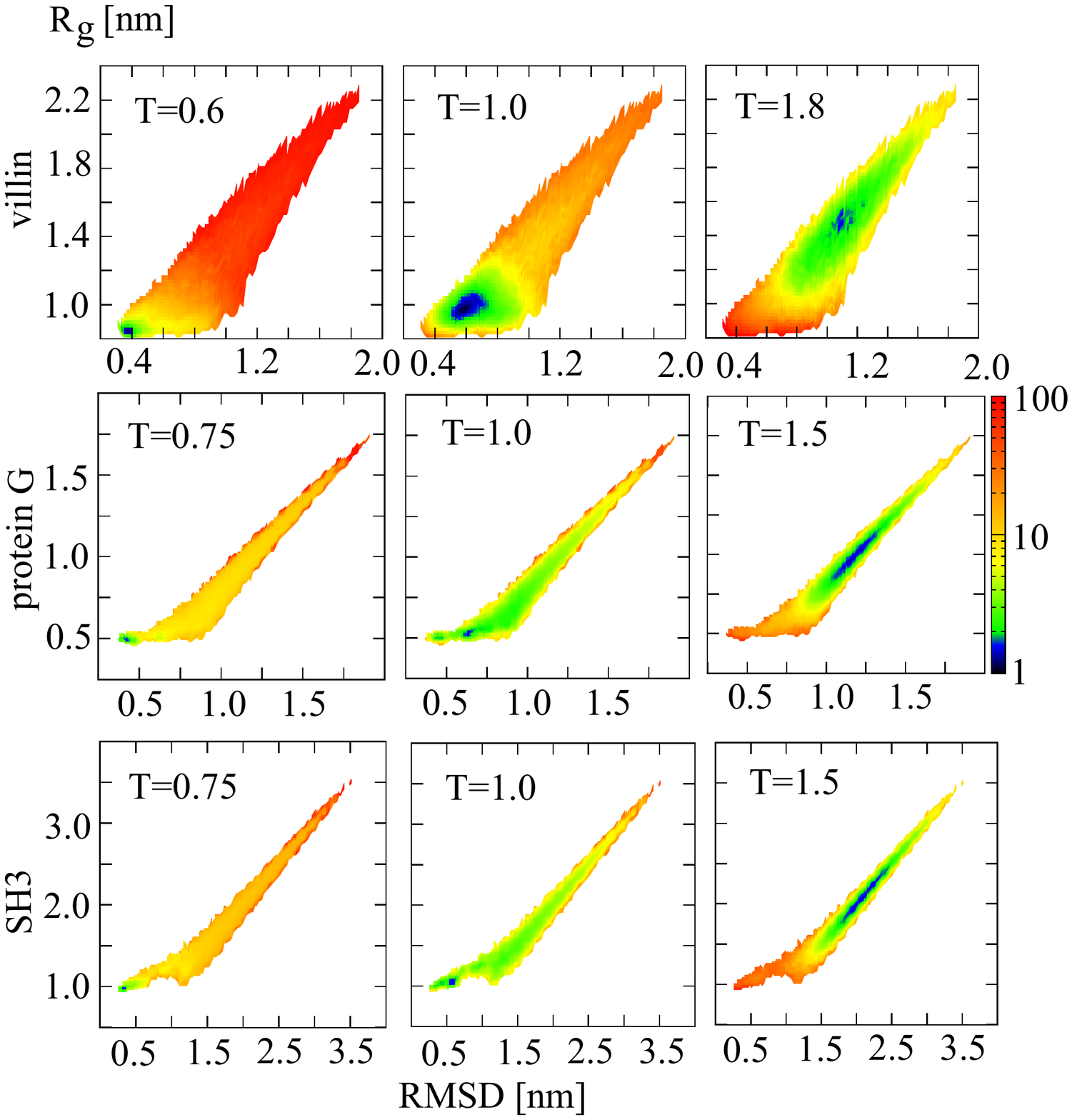}
\caption{The free energy of villin (upper panels), protein G (middle panels) and SH3 (lower panels), caluclated at temperatures below the folding transition (left panels), between the folding and the coil--globule transition (middle panels) and above the coil--globule transition (left panels), as a function of RMSD and gyration radius.}
\label{fig:free}
\end{figure}

Moreover, it is interesting to note that the good overall correlation with the experimental data can be obtained only defining the native state using RMSD and $A_W$. Calculating the values of $p_N^{wt}$ and $p_N^{mut}$ as integral over RMSD and gyration radius (cf. Fig. \ref{fig:free}), on RMSD only, or on $A_W$ only give correlations in the range 0.2--0.4. The reason for this difference in the results seems to be that RMSD and $A_W$ are less correlated than RMSD and $R_g$ (cf. Figs. \ref{fig:free} and \ref{fig:freemut}), and consequently are better in defining  the native region. Specifically, the effect of mutations increase the probability of conformations with values of $A_W$ larger than that of the wild--type protein, mantaining a rather small RMSD, as shown in Fig. \ref{fig:freemut}.

 \section{Frustrated atomic contacts in the native conformation}
 
\begin{figure}
\includegraphics[width=5cm]{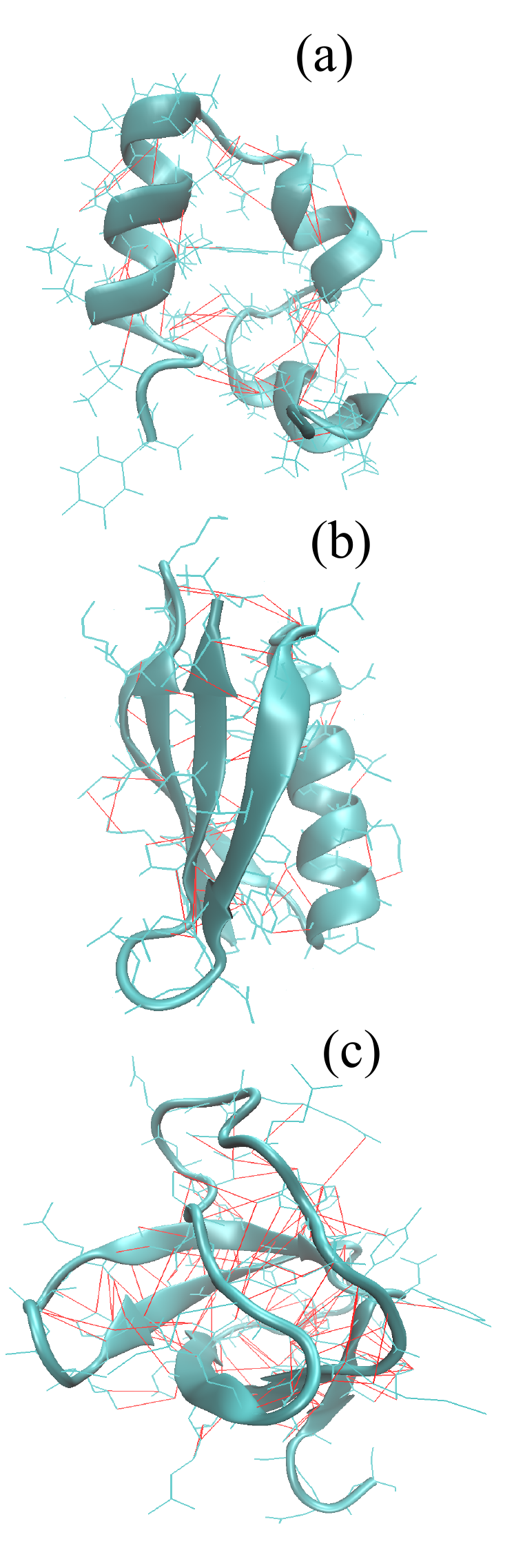}
\caption{The native structure of (a) villin, (b) protein G and (c) SH3 with the frustrated ($B_{\sigma\pi}>0$) atom--atom contacts highlighted in red.}
\label{fig:frust}
\end{figure}

A nice feature of the potential developed above is that, being defined with respect to atom types (and not on atom identifiers, like in Go models),  it include some degree of frustration, which is known to be present in proteins\cite{Bryngelson:1987uu}. One can thus inspect the energy map of the native conformations of the three proteins already discussed, to identify repulsive contacts,  defined as those displaying $B_{\sigma\pi}>0$. Such contacts are highlighted in red in Fig. \ref{fig:frust}.
 
There are 11.9\% frustrated contacts in villin, 8.6\% in protein G and 20.3\% in SH3, numbers that are comparable to those found in similar calculations carried out with other potentials\cite{Lui:2013ec,Jenik:2012jo}. In the case of villin, they are localized mainly in the third helix and in the tertiary contacts between the first helix and the other two. This agrees qualitatively with the result of a similar investigations carried out with the help of an evolutionary--derived potential\cite{Lui:2013ec} and of an associative--memory potential\cite{Jenik:2012jo}, which emphasise the frustration of contacts within the third helix and between the first and the second helix. In the case of protein G, the present model identifies frustrated contacts in the helix and in the terminal part of the first hairpin, while the associative--memory potential in the helix and in the secon hairpin. In the case of SH3, the optimized potential reveals frustrated contacts in the terminal beta-sheet, between the RT loop and the distal hairpin and in the stem of the distal hairpin, while the associative--memory potential in the stem of distal hairpin and in the stem of of the RT loop and the evolutionary--derived potential in the stem of the distal hairpin, in the RT loop, between these two and in the n--src loop.
 
The small differences observed in the frustration maps generated in present and in other works are most probably due to the fact that our potential is atom-based, while the others are amino acid-based. This means that repulsive and attractive interaction between pairs of atoms between two given amino acids, as  predicted by the present model, can sum together to give a total interaction which can be either repulsive or attractive. Consequently, the present model provide an information which is complementary to that of the other two.
 
 \section{Conclusions}
 
 In spite of the continuously growing capability of algorithms and computers to perform longer simulations of larger systems with portable, explicit--solvent potentials, implicit--solvent models of biomolecules interacting with simplified potentials can still be useful for many applications, like very-large systems, mutation scans and aggregation studies. So far, this kind of problems were tackled making use of Go models, which neglects the residual frustration present in all proteins. The model discussed in the present work is based on an optimization of the matrix which controls the interaction between atom types to make the experimental native conformation as the low--temperature equilibrium state of the system. This model can reproduce a number of known data about proteins, like the stability of their native state, the two--state transition, the energetic effect of mutations on their stability, while still displaying a realistic degree of frustration.
 
We think that the strength of this approach is its versatility.  One can use as input for the optimization of the potential any set of experimental data, even an heterogeneous one, provided that they can be expressed as thermal averages of some conformational property. For example, we showed that the structure of the denatured state of the proteins used in the present work is not in complete agreement with the NMR data in denaturing conditions. This is not really unexpected, since the input data we used describe the native conformation, and consequently the predictions of the model cannot but worsen as they involve states which are distant from the native state. To improve the model, one can thus introduce in the optimization data concerning the denatured state. Moreover, this approach can be used to correct existing potentials, even in explicit solvent, for specific goals. It is enough to use the potential to be corrected as initial potential of the optimization procedure. 
 

\newpage

\end{document}